\documentclass[aps,prc,preprint,showpacs,superscriptaddress]{revtex4-1}
\usepackage[english]{babel}
\usepackage[T1]{fontenc}
\usepackage{amssymb}
\usepackage{amsmath}
\usepackage{graphicx}
\usepackage{epstopdf}
\usepackage{geometry}
\usepackage{bm}
\DeclareMathOperator\erf{erf}
\begin{document}
\renewcommand{\eqref}[1]{(\ref{#1})}
\newcommand{\Eq}[1]{Eq.~(\ref{#1})}
\newcommand{\Eqs}[1]{Eqs.~(\ref{#1})}
\newcommand{\Sec}[1]{Sec.~\ref{#1}}
\newcommand{\Ref}[1]{Ref.~\cite{#1}}
\newcommand{\Refs}[1]{Refs.~\cite{#1}}
\newcommand{\Fig}[1]{Fig.~\ref{#1}}
\newcommand{\Figs}[1]{Figs.~\ref{#1}}
\newcommand {\beq} {\begin{eqnarray}}
\newcommand {\eeq} {\end{eqnarray}}
\newcommand {\eeqn} [1] {\label{#1} \end{eqnarray}}
\newcommand {\eoln} [1] {\label{#1} \\}
\newcommand {\eol} {\nonumber \\}
\newcommand {\etal} {\mbox{\it et al}}
\title{Comparison of potential models of nucleus-nucleus bremsstrahlung}

\author{J. Dohet-Eraly}
\email[]{jdoheter@triumf.ca}
\affiliation{TRIUMF, 4004 Wesbrook Mall, Vancouver, BC V6T 2A3,Canada}
\affiliation{Physique Quantique, C. P. 165/82, and\\
Physique Nucl\'eaire Th\'eorique et Physique Math\'ematique, C.P. 229, Universit\'e Libre de Bruxelles (ULB), B-1050 Brussels Belgium.}
\author{D. Baye}
\email[]{dbaye@ulb.ac.be}
\affiliation{Physique Quantique, C. P. 165/82, and\\
Physique Nucl\'eaire Th\'eorique et Physique Math\'ematique, C.P. 229, Universit\'e Libre de Bruxelles (ULB), B-1050 Brussels Belgium.}

\date{\today}

\begin{abstract}
At low photon energies, the potential models of nucleus-nucleus bremsstrahlung are based on electric transition multipole operators, which are derived either only from the nuclear current or only from the charge density by making the long-wavelength approximation and using the Siegert theorem. 
In the latter case, the bremsstrahlung matrix elements are divergent and some regularization techniques are used to obtain finite values for the bremsstrahlung cross sections. 
From an extension of the Siegert theorem, which is not based on the long-wavelength approximation, a new potential model of nucleus-nucleus bremsstrahlung is developed. Only convergent integrals are included in this approach.
Formal links between bremsstrahlung cross sections obtained in these different models are made.
Furthermore, three different ways to calculate the regularized matrix elements are discussed and criticized.
Some prescriptions for a proper implementation of the regularization are deduced.
A numerical comparison between the different models is done by applying them to the $\alpha+\alpha$ bremsstrahlung.  
\end{abstract}
\maketitle
\section{Introduction}
Nuclear bremsstrahlung refers to a radiative transition between nuclear states which lie in the continuum.
This paper principally focuses on nucleus-nucleus bremsstrahlung, where the photon emission is induced by a collision between two nuclei or a nucleus and a neutron. 
However, the emission of bremsstrahlung photons can also accompany proton decays, $\alpha$ decays, or fissions. 
The common essential feature of these processes is that both initial and final states are not square-integrable in stationary approaches.
This feature leads in some bremsstrahlung models~\cite{TM85,LR86c,GJF12,GJF14} to divergent matrix elements, which have to be replaced by some finite values via some regularization prescription.
Then, the difficult problem of analyzing the influence of the regularization techniques on the results arises.
This problem is avoided in other bremsstrahlung models~\cite{PH82,BD85,La86,LR86,LTK90a,LTK90b,BSD91,BDK92,LTK92, Ma11, Ma12}, which are based from the beginning only on convergent matrix elements.
To understand the presence or the absence of divergence problems in different bremsstrahlung models, it is required to discuss the fundamental bases of these models. 
This discussion is also useful to highlight the links between these models.

The description of electromagnetic transitions in nuclear systems relies on the interaction between the nuclear current and the electromagnetic field. 
When the long-wavelength approximation (LWA) can be applied, the interaction between the nuclear current and the electric field does not explicitly depend on the nuclear current anymore but can be deduced exclusively from the charge density.
This property is referred to as the Siegert theorem~\cite{Si37}.
This is particularly useful in nuclear physics where the current density is usually less well known than the charge density.
However, in the study of radiative transitions between continuum states, the long-wavelength approximation leads to mathematical divergences and the dependence on the nuclear current cannot thus be fully removed in bremsstrahlung models.

To avoid this divergence problem, most authors decided not to apply the Siegert theorem in bremsstrahlung models~\cite{PH82,BD85,La86, LR86,LTK90a,LTK90b,BSD91,BDK92,LTK92, Ma11, Ma12}. 
For potential models of bremsstrahlung, where the colliding nuclei are treated as point-like particles interacting with an effective nucleus-nucleus interaction, some authors preferred to apply the Siegert theorem and to replace the divergent integrals by convergent expressions by using some regularization techniques~\cite{TM85,LR86c,GJF12,GJF14}.
Even if applying the Siegert theorem seems to simplify the expressions of the matrix elements required to evaluate the bremsstrahlung cross sections, the regularization techniques used in~\Refs{TM85,LR86c,GJF12,GJF14} break this apparent simplicity.

In a recent paper~\cite{DEB13}, an extension of the Siegert theorem~\cite{SWA90}, which does not rely to the long-wavelength approximation and which does not lead to divergent matrix elements, was proposed to greatly reduce the dependence of the electric transition multipole operators on the nuclear current. 
This method was applied to a microscopic description of nucleus-nucleus bremsstrahlung, namely the $\alpha+\alpha$~\cite{DEB13} and $\alpha+N$ systems~\cite{DE14}.
In this paper, the method developed in~\Ref{DEB13} is applied to a potential model of bremsstrahlung.
With this method, the expressions of bremsstrahlung cross sections obtained after regularization in the Siegert approach based on the long-wavelength approximation can be justified without introducing divergent integrals. 

In \Sec{Seccross}, the potential models of bremsstrahlung are outlined. 
In \Sec{Secelec}, the different forms of the electric transition multipole operators are derived in a common framework. 
The interest of a Siegert approach in the potential models of bremsstrahlung is discussed.
In \Sec{Secmatel}, the calculation of the matrix elements of the electric transition multipole operators is explained and the basic idea of the regularization techniques is presented. 
In \Sec{Secreg}, three implementations of regularization techniques are presented and compared: the fixed $\epsilon_0$ method proposed by Garrido, Fedorov, and Jensen in~\Ref{GJF12}, the integration by parts (IP) method inspired by Tanimura and Mosel's work~\cite{TM85}, and the contour integration (CI) method, more adapted for numerical calculations, based on the contour integration proposed by Vincent and Fortune~\cite{VF70}.
In \Sec{Secres}, the different versions of the potential model of nucleus-nucleus bremsstrahlung are applied to the $\alpha+\alpha$ system and the bremsstrahlung cross sections are compared.
Concluding remarks are presented in \Sec{Secconcl}.
\section{Bremsstrahlung cross sections}\label{Seccross}
In the center-of-mass (c.m.)\ frame, two spinless nuclei with charges $Z_1 e$ and $Z_2 e$, masses $m_1$ and $m_2$, respectively, and reduced mass $\mu_M$ collide with initial relative wave vector $\mathbf{k}_i$ in the $z$ direction and relative energy $E_i=\hbar^2 k^2_i/2\mu_M$. 
After emission of a photon in direction $\Omega_\gamma$ with energy $E_\gamma=\hbar k_\gamma c$, the nuclear system has final relative vector $\mathbf{k}_f$ in direction $\Omega_f=(\theta_f,\varphi_f)$ and relative energy $E_f=\hbar^2 k^2_f/2\mu_M$ given by
\begin{equation}\label{consen}
E_f=E_i-E_\gamma,
\end{equation}
up to small recoil corrections.

The bremsstrahlung cross sections are evaluated from the multipole matrix elements, which are proportional to the matrix elements of the electromagnetic transition multipole operators $\mathcal{M}^\sigma_{\lambda\mu}$ 
between the incoming initial state $\Psi^+_i$ in the $z$ direction with energy $E_i$ and the outgoing final state $\Psi^-_f(\Omega_f)$ in direction $\Omega_f$ with energy $E_f$,
\begin{equation}
u^\sigma_{\lambda\mu}(\Omega_f)=\alpha^\sigma_\lambda \langle\Psi^-_f(\Omega_f)|\mathcal{M}^\sigma_{\lambda\mu}|\Psi^+_i\rangle,
\end{equation}
where $\lambda$ is the order of the multipole, $\mu$ is its component, $\sigma=0$ or $E$ corresponds to an electric multipole and $\sigma=1$ or $M$ corresponds to a magnetic multipole, and $\alpha^\sigma_\lambda$ is given by 
\begin{equation}
\alpha^\sigma_\lambda=-\frac{\sqrt{2\pi(\lambda+1)}i^{\lambda+\sigma}k^\lambda_\gamma}{\sqrt{\lambda(2\lambda+1)}(2\lambda-1)!!}.
\end{equation}
The differential bremsstrahlung cross section $d\sigma/d E_\gamma$ is given by~\cite{BSD91}
\begin{equation}\label{dsigmadE}
\frac{{\rm d}\sigma}{{\rm d}E_\gamma}=
\frac{E_\gamma}{\pi^2\hbar^5 c }\frac{p^2_f}{1+\delta_{12}} \sum_{\sigma\lambda\mu}
\int^\pi_0 \frac{|u^{\sigma}_{\lambda\mu}(\theta_{f},0)|^2}{2\lambda+1} \sin\theta_f {\rm d}\theta_f,
\end{equation}
where $\delta_{12}$ is equal to unity if nuclei $1$ and $2$ are identical and to zero otherwise. 
The division by $(1+\delta_{12})$ is added to take the possible identity of both nuclei into account.
Other differential bremsstrahlung cross sections are also obtained from the multipole matrix elements $u^{\sigma}_{\lambda\mu}$. 
Explicit formulas can be found in~\Ref{BSD91}.

In the potential model, nuclei are treated as point-like particles interacting with an effective nucleus-nucleus interaction.
The initial and final states $\Psi^+_i$ and $\Psi^-_f$ are solutions of the Schr\"odinger equation
\begin{equation}
H \Psi=E \Psi
\end{equation}
with energy $E_i$ and $E_f$, respectively. 
The internal Hamiltonian $H$ reads
\begin{equation}\label{defH}
H=-\frac{\hbar^2}{2\mu_M}\Delta_{\bm{\rho}}+V(\rho),
\end{equation}
where $\bm{\rho}$ is the relative coordinate between the nuclei, $\rho$ is the norm of $\bm{\rho}$, and $V$ is a local potential describing the interaction between both nuclei. 
The potential $V$ is assumed to be real and central. 
Some comments about more general potentials are given in \Sec{Secelec}.
The potential $U$ can be defined by subtracting the bare Coulomb potential from the potential $V$, 
\begin{equation}
U(\rho)=V(\rho)-\frac{Z_1 Z_2 e^2}{\rho}.
\end{equation}
It is assumed to have a finite range.

The initial and final states $\Psi^+_i$ and $\Psi^-_f(\Omega_f)$ can be expanded in partial-wave series~\cite{BSD91}
\begin{eqnarray}\label{partw1}
\Psi^+_i &=& \sum_{l_i} \psi^i_{l_i 0},\\ \label{partw2}
\Psi^-_f(\Omega_f) &=& 2\sqrt{\pi}\sum_{l_f m_f} (2 l_f+1)^{-1/2} Y^{m_f *}_{l_f}(\Omega_f) e^{-2 i(\sigma_{l_f}+\delta_{l_f})} \psi^f_{l_f m_f},
\end{eqnarray}
where $\sigma_{l_f}$ and $\delta_{l_f}$ are the Coulomb and quasinuclear phase shifts.
Their dependence on energy is dropped to simplify the notation. 
The normalized spherical harmonics $Y^{m_f}_{l_f}$ are defined by following the Condon and Shortley convention.
If colliding nuclei are identical bosons (resp. fermions), partial-wave expansions \eqref{partw1} and \eqref{partw2} are restricted to even (resp. odd) values of $l_i$ and $l_f$ to satisfy the Pauli principle.

The partial waves can be written, by splitting the radial and angular dependences, as
\begin{equation}
\psi^{c}_{l_c m_c}(\bm{\rho})=C_{l_c} \frac{u^{c}_{l_c}(\rho)}{\rho} Y^{m_c}_{l_c}(\Omega_\rho),
\end{equation}
where $\Omega_\rho$ is the angular part of the spherical coordinates of $\bm{\rho}$, $c=i$ or $f$ designates the initial or final channel, and $C_{l_c}$ is a complex coefficient defined by
\begin{equation}
C_{l_c}=2 \sqrt{\pi}(2l_c+1)^{1/2} i^{l_c} e^{i(\sigma_{l_c}+\delta_{l_c})}.
\end{equation}
The radial function $u^c_{l_c}$ is a real solution of the radial Schr\"odinger equation at energy $E_c$,
\begin{equation}\label{Eqrad}
-\beta u^{c\prime\prime}_{l_c}(\rho) + \left[ V_{l_c}(\rho)+V(\rho) \right] u^{c}_{l_c}(\rho)=
E_c u^{c}_{l_c}(\rho),
\end{equation}
where the prime designates the derivative with respect to $\rho$, $\beta=\hbar^2/2\mu_M$, and $V_{l_c}$ is the centrifugal potential
\begin{equation}
V_{l_c}(\rho)=\beta \frac{l_c (l_c+1)}{\rho^2}. 
\end{equation}
The normalization of $u^c_{l_c}(\rho)$ is fixed by its asymptotic behavior,
\begin{eqnarray}\label{as1}
u^{c}_{l_c}(\rho)\mathop{\longrightarrow}\limits_{\rho\to \infty}u^{c,{\rm as}}_{l_c}(\rho) &=& \frac{(1+\delta_{12})^{1/2}}{\sqrt{v_c} k_c}
\left[F_{l_c}(\eta_c,k_c\rho)\cos\delta_{l_c}+G_{l_c}(\eta_c,k_c\rho)\sin\delta_{l_c}\right]\\ \label{as2}
&=&i \frac{(1+\delta_{12})^{1/2}}{2 \sqrt{v_c} k_c}
\left[e^{-i\delta_{l_c}}I_{l_c}(\eta_c,k_c\rho)-e^{i\delta_{l_c}}O_{l_c}(\eta_c,k_c\rho)\right],
\end{eqnarray}
where $\eta_c$ is the Sommerfeld parameter, $F_{l_c}$ and $G_{l_c}$ are the regular and irregular Coulomb functions, and $I_{l_c}$ and $O_{l_c}$ are the incoming and outgoing Coulomb wave functions.
From \Eqs{partw1} and \eqref{partw2}, the expansion of $u^\sigma_{\lambda\mu}$ in partial-wave series can be written as~\cite{BSD91}
\begin{equation}\label{upart}
u^\sigma_{\lambda\mu}=2\sqrt{\pi}\alpha^\sigma_\lambda \sum_{l_i l_f} (2l_f+1)^{-1/2} Y^\mu_{l_f} (\Omega_f) (l_i \lambda 0 \mu| l_f \mu) e^{2 i(\sigma_{l_f}+\delta_{l_f})}
\langle\psi^{f}_{l_f}||\mathcal{M}^\sigma_\lambda|| \psi^{i}_{l_i}\rangle,
\end{equation} 
where the reduced matrix elements are defined following the convention
\begin{equation}
\langle\psi^f_{l_f m_f}|\mathcal{M}^\sigma_{\lambda\mu}| \psi^{i}_{l_i m_i}\rangle=(l_i \lambda m_i \mu|l_f m_f) \langle\psi^{f}_{l_f}||\mathcal{M}^\sigma_\lambda|| \psi^{i}_{l_i}\rangle.
\end{equation} 
Since only the electric transitions ($\sigma=E$) are concerned by the Siegert approach and since they dominate for light-ion bremsstrahlung at low photon energy~\cite{PH82}, the magnetic transitions are not considered hereafter.
The potential models of nucleus-nucleus bremsstrahlung used in~\Refs{BSD91,TM85,LR86,La86,LR86c,GJF12,GJF14} differ by their definitions of the electric transition multipole operators, which are given in the next section.
\section{Electric transition multipole operators}\label{Secelec}
The electric transition multipole operators can be defined from the nuclear current by~\cite{BM69}
\begin{equation}\label{defME}
\mathcal{M}^{E}_{\lambda\mu}=\sqrt{\frac{\lambda}{\lambda+1}}\frac{(2\lambda+1)!!}{k^\lambda_\gamma c} \int\mathbf{J}(\mathbf{r})\cdot\mathbf{A}^{E}_{\lambda\mu}(\mathbf{r}) d\mathbf{r},
\end{equation}
where $\mathbf{J}$ is the nuclear current density and $\mathbf{A}^{E}_{\lambda\mu}$ is the electric multipole defined, in the Coulomb gauge, as~\cite{Me62}
\begin{equation}\label{defAE}
\mathbf{A}^{E}_{\lambda\mu}(\mathbf{r})=\frac{i}{k_\gamma \sqrt{\lambda (\lambda+1)}} \bm{\chi}_{\lambda\mu}(k_\gamma,\mathbf{r})
\end{equation}
with
\begin{equation}
\bm{\chi}_{\lambda\mu}(k,\mathbf{r})=\left(k^2 \mathbf{r}+\bm{\nabla} \frac{\partial}{\partial r}r \right)j_\lambda(k r)Y^\mu_{\lambda}(\Omega)
\end{equation}
and $\mathbf{r}=(r,\Omega)$.
The usual notation $j_\lambda$ is used to designate the spherical Bessel functions (first kind) of order $\lambda$~\cite{AS65}.

The suppression of the current dependence of the electric transitions at low photon energies relies on the fact that $\mathbf{A}^{E}_{\lambda\mu}$ is reduced to a gradient term at the long-wavelength approximation, i.e.,\ by keeping only the lowest order term in $k_\gamma r$ in the expression of the electric multipole,
\begin{equation}
\mathbf{A}^{E}_{\lambda\mu}(\mathbf{r})\mathop{\longrightarrow}\limits_{k_\gamma\to 0}
 \frac{i\sqrt{\lambda+1}k_\gamma^{\lambda-1}}{\sqrt{\lambda}(2\lambda+1)!!}\bm{\nabla} r^\lambda Y^\mu_\lambda(\Omega).
\end{equation}
To reduce the current dependence without applying the long-wavelength approximation, the idea is to introduce an approximate electric transition multipole operator, denoted by $\widetilde{\mathcal{M}}^{E}_{\lambda\mu}$, in which $\mathbf{A}^{E}_{\lambda\mu}$ is approximated only by a gradient term
\begin{equation}\label{defMEapp}
\widetilde{\mathcal{M}}^{E}_{\lambda\mu}=\sqrt{\frac{\lambda}{\lambda+1}}\frac{(2\lambda+1)!!}{k^\lambda_\gamma c} \int\mathbf{J}(\mathbf{r})\cdot\bm{\nabla} \Phi_{\lambda\mu}(\mathbf{r}) d\mathbf{r},
\end{equation}
where $\Phi_{\lambda\mu}$ is chosen such that $\bm{\nabla}\Phi_{\lambda\mu}$ and $\mathbf{A}^{E}_{\lambda\mu}$ have the same behavior at low photon energies,
\begin{equation}
\frac{||\bm{\nabla}\Phi_{\lambda\mu}||}{||\mathbf{A}^{E}_{\lambda\mu}||}\mathop{\longrightarrow}\limits_{k_\gamma\to 0}1.
\end{equation}
Practical choices of $\Phi_{\lambda\mu}$ are specified in \Eqs{choicephi} and \eqref{choicephi2}.

After integrating by parts and by using the continuity equation
\begin{equation}\label{currentcons}
\bm{\nabla}\cdot \mathbf{J}(\mathbf{r})+\frac{i}{\hbar}\left[H,\rho(\mathbf{r})\right]=0,
\end{equation}
where $\rho(\mathbf{r})$ is the charge density, the operator $\widetilde{\mathcal{M}}^{E}_{\lambda\mu}$ can be written as
\begin{equation}
\widetilde{\mathcal{M}}^{E}_{\lambda\mu}=i \sqrt{\frac{\lambda}{\lambda+1}}\frac{(2\lambda+1)!!}{k^\lambda_\gamma \hbar c}\int \left[H,\rho(\mathbf{r})\right] \Phi_{\lambda\mu}(\mathbf{r}) d\mathbf{r},
\end{equation}
when $\Phi_{\lambda\mu}(\mathbf{r})$ is assumed to lead to a vanishing surface term at infinity.
If the partial waves $\psi^{i}_{l_i m_i}$ and $\psi^{f}_{l_f m_f}$ are assumed to be exact eigenstates of the Hamiltonian defined by \Eq{defH}, the matrix elements of the approximate electric transition multipole operators between initial and final states are given by 
\begin{equation}\label{Siegert}
\langle \psi^{f}_{l_f m_f} | \widetilde{\mathcal{M}}^{E}_{\lambda\mu} |\psi^{i}_{l_i m_i}\rangle = -i\sqrt{\frac{\lambda}{\lambda+1}} \frac{(2\lambda+1)!!}{k^{\lambda-1}_\gamma }\int \langle \psi^{f}_{l_f m_f} | \rho(\mathbf{r}) \Phi_{\lambda\mu}(\mathbf{r})| \psi^{i}_{l_i m_i}\rangle d\mathbf{r},
\end{equation}
where \Eq{consen} is used.
The r.h.s.\ of \Eq{Siegert} defines the Siegert form of the approximate electric transition multipole operator, 
denoted as $\widetilde{\mathcal{M}}^{E{\rm (S)}}_{\lambda\mu}$, which depends on the charge density and not on the current density,
\begin{equation}\label{Siegert2}
\widetilde{\mathcal{M}}^{E{\rm (S)}}_{\lambda\mu} = -i \sqrt{\frac{\lambda}{\lambda+1}} \frac{(2\lambda+1)!!}{k_\gamma^{\lambda-1}}
\int \rho(\mathbf{r}) \Phi_{\lambda\mu}(\mathbf{r}) d\mathbf{r}.
\end{equation}
The electric transition multipole operator $\mathcal{M}^{E}_{\lambda\mu}$ can be written from $\widetilde{\mathcal{M}}^{E}_{\lambda\mu}$ by adding a correcting term,
\begin{equation}\label{Mcorr}
\mathcal{M}^{E}_{\lambda\mu}=\widetilde{\mathcal{M}}^{E}_{\lambda\mu}+(\mathcal{M}^{E}_{\lambda\mu}-\widetilde{\mathcal{M}}^{E}_{\lambda\mu}).
\end{equation}
At low photon energies, the contribution of the correcting term $\mathcal{M}^{E}_{\lambda\mu}-\widetilde{\mathcal{M}}^{E}_{\lambda\mu}$ should be weak compared to the contribution of $\widetilde{\mathcal{M}}^{E}_{\lambda\mu}$.
By analogy with \Eq{Mcorr}, the Siegert form of the electric transition multipole operator, denoted by $\mathcal{M}^{E{\rm (S)}}_{\lambda\mu}$, can be defined by~\cite{SWA90,DEB13}
\begin{equation}\label{MESiegert}
\mathcal{M}^{E{\rm (S)}}_{\lambda\mu}=\widetilde{\mathcal{M}}^{E{\rm (S)}}_{\lambda\mu}+(\mathcal{M}^{E}_{\lambda\mu}-\widetilde{\mathcal{M}}^{E}_{\lambda\mu}).
\end{equation}
Since at low photon energies the contribution of $\widetilde{\mathcal{M}}^{E{\rm (S)}}_{\lambda\mu}$, which is current-independent, dominates, the current dependence is well reduced in the Siegert operator $\mathcal{M}^{E{\rm (S)}}_{\lambda\mu}$ in comparison with the non-Siegert operator $\mathcal{M}^{E}_{\lambda\mu}$.
The non-Siegert and Siegert operators, defined by \Eqs{defME} and \eqref{MESiegert}, {\it exactly} lead to the same results if consistent current and charge densities are considered and the exact eigenstates of Hamiltonian \eqref{defH} are used.
Consequently, the non-uniqueness of $\widetilde{\mathcal{M}}^{E}_{\lambda\mu}$ or equivalently the arbitrary nature of the choice of $\Phi_{\lambda\mu}$ is not problematic since it has theoretically no influence.

Possible choices of $\Phi_{\lambda\mu}$, which avoid divergent integrals in bremsstrahlung calculations, are given by
\begin{equation}\label{choicephi}
\Phi_{\lambda\mu}(\mathbf{r})=\frac{i\sqrt{\lambda+1}}{k_\gamma \sqrt{\lambda}} \varphi_\lambda (k_\gamma r,\epsilon) Y^\mu_\lambda(\Omega),
\end{equation}
where
\begin{equation}\label{choicephi2}
\varphi_\lambda(x,\epsilon ) = j_\lambda(x)  {\rm \ or \ } \frac{x^\lambda}{(2\lambda+1)!!} e^{-\epsilon x} {\rm \ or \ }
 \frac{x^\lambda}{(2\lambda+1)!!} e^{-\epsilon^2 x^2}
\end{equation}
with $0<\epsilon\ll 1$.
These choices are named respectively the Bessel, exponential, and Gaussian choices.
The parameter $\epsilon$ has no meaning for the Bessel choice but is denoted for having a common notation. 
The Bessel choice is used in~\Refs{DEB13,DE14}. 
The exponential and Gaussian choices are used in the next section to make some formal link between the results based on the extended Siegert theorem and the ones based on the regularization techniques.

At the long-wavelength approximation, the Siegert operator $\mathcal{M}^{E{\rm (S)}}_{\lambda\mu}$ is reduced to the operator $\mathcal{M}^{E{\rm (S,LWA)}}_{\lambda\mu}$ defined by
\begin{equation}\label{Siegertlw}
\mathcal{M}^{E{\rm (S,LWA)}}_{\lambda\mu} = \int \rho(\mathbf{r}) r^\lambda Y^\mu_\lambda(\Omega)  d\mathbf{r},
\end{equation}
where the current-dependence is fully dropped. 
However, in the time-independent approaches, since the continuum states have an infinite extension, applying the long-wavelength approximation is not rigorously justified in the study of bremsstrahlung.

Let me particularize the electric transition multipole operators to the potential model. 
To limit the complexity of the calculations, the charge and current densities for free nucleons are considered.
For spinless nuclei, the charge and current densities are given by
\begin{equation}\label{rhoext}
\rho(\mathbf{r})=e Z_1 \delta\left(\frac{m_2}{m}\bm{\rho}-\mathbf{r}\right)+e Z_2 \delta\left(\frac{m_1}{m}\bm{\rho}+\mathbf{r}\right)
\end{equation}
and
\begin{equation}\label{Jext}
\mathbf{J}(\mathbf{r})=\frac{e}{2} \left\{\frac{Z_1}{m_1}\left[\mathbf{p}_\rho,\delta\left(\frac{m_2}{m}\bm{\rho}-\mathbf{r}\right)\right]_+-
\frac{Z_2}{m_2}\left[\mathbf{p}_\rho,\delta\left(\frac{m_1}{m}\bm{\rho}+\mathbf{r}\right)\right]_+\right\}.
\end{equation}
The shorthand notation $\left[a,b\right]_+$ is used for $ab+ba$ where $a$ and $b$ can be scalar or vector operators.
For a real central potential, these current and charge densities exactly verify the continuity equation.
Consequently, differences between Siegert and non-Siegert approaches can only come from numerical inaccuracies in the resolution of the radial Schr\"odinger equation or in the computation of the integrals. 
Therefore, choosing the Siegert or non-Siegert approach is a matter of convenience and should have no significant impact on the results. 

Let me note that the continuity equation \eqref{currentcons} is generally not verified if the nucleus-nucleus potential is not purely central. 
For instance, if the potential contains some parity-dependent terms, an extra current should be considered to verify \Eq{currentcons}.
Similarly, if the spins of the colliding nuclei are considered and if the potential contains a spin-orbit term, a spin-orbit contribution should be added to the nuclear current for verifying \Eq{currentcons}.
In both cases, neglecting these extra currents should have a smaller importance in the Siegert approach than in the non-Siegert approach, especially at low-photon energy. 
If only the convection current, defined by \Eq{Jext} is considered, the Siegert approach should thus be preferred.
In the optical models, the interaction between nuclei is described by a so-called optical potential, i.e.,\ a potential containing an imaginary part which simulates the effects of the open channels not explicitly described. 
For complex potentials, the Hamiltonian is not Hermitian and \Eq{Siegert} is not valid. 
In these models, the non-Siegert electric transition multipole operators have thus to be considered.

Let me restrict again to real central potentials.
Inserting the charge and current densities defined by \Eqs{rhoext} and \eqref{Jext} in \Eq{defME} leads to the explicit definition \eqref{MEnonSdef} of the non-Siegert electric transition multipole operators,
\begin{equation}\label{MEnonSdef}
\mathcal{M}^{E}_{\lambda\mu}=\frac{i e (2\lambda+1)!!}{\mu_M c k^{\lambda+1}_\gamma(\lambda+1)}\left[Z_1 \bm{\chi}_{\lambda\mu}\left(\frac{m_2}{m}k_\gamma,\bm{\rho}\right)+(-1)^\lambda Z_2 \bm{\chi}_{\lambda\mu}\left(\frac{m_1}{m}k_\gamma,\bm{\rho}\right)
\right]\cdot \mathbf{p}_{\rho}
\end{equation}
with $m=m_1+m_2$. 
These non-Siegert electric transition multipole operators are used in several models of bremsstrahlung~\cite{PH82,La86,LR86,BSD91}. 

The approximate non-Siegert and Siegert operators are written in the potential model as
\begin{equation}\label{explappnS}
\widetilde{\mathcal{M}}^{E}_{\lambda\mu}=\frac{i e(2\lambda+1)!!}{2\mu_M c k^{\lambda+1}_\gamma}\left[Z_1 
\bm{\nabla}_\rho \varphi_\lambda\left(\frac{m_2}{m} k_\gamma \rho,\epsilon\right)Y^\mu_\lambda(\Omega_\rho)+
(-1)^\lambda Z_2 \bm{\nabla}_\rho \varphi_\lambda\left(\frac{m_1}{m} k_\gamma \rho,\epsilon\right)Y^\mu_\lambda(\Omega_\rho),\mathbf{p}_\rho\right]_+
\end{equation}
and
\begin{equation}\label{explappS}
\widetilde{\mathcal{M}}^{E {\rm (S)} }_{\lambda\mu} = \frac{e(2\lambda+1)!!}{k^\lambda_\gamma}\left[Z_1 \varphi_\lambda\left(\frac{m_2}{m}k_\gamma \rho,\epsilon\right)+(-1)^\lambda Z_2 \varphi_\lambda\left(\frac{m_1}{m}k_\gamma\rho,\epsilon\right) 
\right]Y^\mu_\lambda(\Omega_\rho).
\end{equation}
The explicit expression of the Siegert electric transition multipole operator in the potential model is obtained from \Eqs{MESiegert}, \eqref{MEnonSdef}, \eqref{explappnS}, and \eqref{explappS}.
Inserting the charge density defined by \Eqs{rhoext} in \Eq{Siegertlw} or applying the long-wavelength approximation to \Eq{explappS} leads to the explicit definition of the long-wavelength-approximated Siegert electric transition multipole operators,
\begin{equation}\label{defMElwpot}
\mathcal{M}^{E{\rm(S,LWA)}}_{\lambda\mu}=e Z^{(\lambda)}_{\rm eff}\rho^\lambda Y^\mu_\lambda(\Omega_\rho),
\end{equation}
where the effective charge $Z^{(\lambda)}_{\rm eff}$ is defined by
\begin{equation}
Z^{(\lambda)}_{\rm eff}= Z_1 \left(\frac{m_2}{m}\right)^\lambda+Z_2\left(-\frac{m_1}{m}\right)^\lambda.
\end{equation}
The LWA Siegert multipole operators are used in~\Refs{TM85,LR86c,GJF12,GJF14}.
Intrinsically, the operator $\mathcal{M}^{E {\rm (S,LWA)} }_{\lambda\mu}$ includes an extra approximation in comparison to the operator $\mathcal{M}^{E}_{\lambda\mu}$. 
However, it has the advantage of having a simpler form which does not include any derivative of the radial wave function.
Nevertheless, this apparent advantage can be lost with some regularization techniques, as in the IP method.
This fact is highlighted in \Sec{Secreg}.

The multipole matrix elements $u^{E}_{\lambda\mu}$ converge by using the electric transition multipole operators $\mathcal{M}^{E}_{\lambda\mu}$, $\widetilde{\mathcal{M}}^E_{\lambda\mu}$, and $\widetilde{\mathcal{M}}^{E {\rm (S)}}_{\lambda\mu}$ whereas they diverge by using the operators $\mathcal{M}^{E{\rm(S,LWA)}}_{\lambda\mu}$.
This property is made apparent in the next section but can already be understood.
Since the wave functions are not square-integrable, the matrix elements $u^{E}_{\lambda\mu}$ converge only if the electric transition multipole operators tend asymptotically to zero, rapidly enough.
Thus, since the operators $\mathcal{M}^{E{\rm(S,LWA)}}_{\lambda\mu}$ are increasing functions of $\rho$, they lead to divergent values of the multipole matrix elements $u^E_{\lambda\mu}$.
For discussing the asymptotic behavior of $\mathcal{M}^{E}_{\lambda\mu}$, the scalar product $\bm{\chi}_{\lambda\mu}(k,\bm{\rho})\cdot \mathbf{p}_{\rho}$ is advantageously written as
\begin{equation}\label{chip}
\bm{\chi}_{\lambda\mu}(k,\bm{\rho})\cdot \mathbf{p}_{\rho}=-i  \frac{\hbar \lambda(\lambda+1)}{\rho}j_\lambda(k\rho)Y^\mu_\lambda(\Omega_\rho) \frac{\partial}{\partial \rho}
-i  \frac{\hbar}{\rho^2}\left[\frac{\partial}{\partial \rho}\rho j_\lambda(k\rho)\right]
\left[\bm{\nabla}_{\Omega_\rho} Y^\mu_\lambda(\Omega_\rho) \right]\cdot \bm{\nabla}_{\Omega_\rho},
\end{equation}
which can be deduced from the properties of the spherical Bessel functions~\cite{AS65}. 
The angular operator $\bm{\nabla}_{\Omega_\rho}$ is implicitly defined by~\cite{VMK88}
\begin{equation}
\bm{\nabla}_{\rho}=\frac{\bm{\rho}}{\rho} \frac{\partial}{\partial \rho} +\frac{1}{\rho} \bm{\nabla}_{\Omega_\rho}.
\end{equation}
Since the spherical Bessel functions $j_\lambda(k\rho)$ behave asymptotically as oscillating functions divided by $\rho$~\cite{AS65}, \Eq{chip} shows that the electric transition multipole operators $\mathcal{M}^{E}_{\lambda\mu}$ behave asymptotically as oscillating functions divided by $\rho^2$.
The radial wave functions and thus the partial waves behave asymptotically as oscillating functions, as it can be seen from \Eq{as1} or \eqref{as2}.
By combining both these properties, the matrix elements $\langle\psi^{f}_{l_f}||\mathcal{M}^{E}_{\lambda}|| \psi^{i}_{l_i}\rangle$ and thus the matrix elements $u^\sigma_{\lambda\mu}$ are proved to be convergent. 
It can be shown by a similar reasoning that $\widetilde{\mathcal{M}}^E_{\lambda\mu}$ and $\widetilde{\mathcal{M}}^{E {\rm (S)}}_{\lambda\mu}$ also lead to convergent matrix elements $u^\sigma_{\lambda\mu}$.
\section{Matrix elements of the electric transition multipole operators between partial waves}\label{Secmatel}
The reduced matrix elements of the non-Siegert multipole operators $\mathcal{M}^{E}_{\lambda\mu}$ between partial waves are given by~\cite{BD85}
\begin{equation}\label{ME}
\langle \psi^f_{l_f}||\mathcal{M}^{E}_{\lambda}|| \psi^i_{l_i}\rangle=
 \frac{(2\lambda+1)!!}{k^{\lambda+1}_\gamma}\frac{ e\hbar}{\mu_M c}\mathbb{Y}_{l_f\lambda l_i} C^*_{l_f}C_{l_i}
\left[Z_1 \mathcal{I}_\lambda \left(\frac{m_2}{m} k_\gamma\right)+(-1)^\lambda Z_2 \mathcal{I}_\lambda\left(\frac{m_1}{m} k_\gamma\right)\right],
\end{equation}
where $\mathbb{Y}_{l_f\lambda l_i}$ is a shorthand notation for the following reduced matrix element~\cite{Ed57}
\begin{equation}
\mathbb{Y}_{l_f\lambda l_i}=\langle Y_{l_f} || Y_\lambda || Y_{l_i} \rangle=(-1)^\lambda (4\pi)^{-1/2} (2\lambda+1)^{1/2} (l_f \lambda 00|l_i0)
\end{equation}
and where $\mathcal{I}_\lambda$ is given by
\begin{equation}\label{IE}
\begin{split}
\mathcal{I}_\lambda (k)=&\frac{\lambda(\lambda+1)+l_i(l_i+1)-l_f(l_f+1)}{2(\lambda+1)}\int^\infty_0 \frac{u^f_{l_f} u^i_{l_i}}{\rho^2} \left[\rho j_\lambda(k\rho)\right]'  d\rho\\ 
&+\lambda \int^\infty_0 u^f_{l_f} j_\lambda(k\rho) (u^i_{l_i}/\rho)' d\rho.
\end{split}
\end{equation}
The dependence on $\rho$ of the radial functions $u^f_{l_f}$ and $u^i_{l_i}$ is dropped to simplify the notations. 
For continuum to continuum transitions, the integrands in \Eqs{IE} behave asymptotically as oscillating functions divided by $\rho^2$, as anticipated in \Sec{Seccross}. 
The integrals thus converge but slowly.
The convergence rate can be improved by using the contour integration method proposed in~\Ref{VF70} and largely used in bremsstrahlung models~\cite{PH82,BD85,La86,LR86,LTK90a,DEB13,DE14}. 
The principle of this method is explained in \Sec{subSeccontour}.

The reduced matrix elements of the approximate multipole operators between partial waves are given in the Siegert approach by
\begin{equation}\label{MES}
\langle \psi^f_{l_f}||\widetilde{\mathcal{M}}^{E{\rm (S)}}_{\lambda}|| \psi^i_{l_i}\rangle=
 \frac{(2\lambda+1)!!e}{k^{\lambda}_\gamma}\mathbb{Y}_{l_f\lambda l_i} C^*_{l_f}C_{l_i}
\left[Z_1 \widetilde{\mathcal{I}}^{\rm (S)}_\lambda \left(\frac{m_2}{m} k_\gamma\right)+(-1)^\lambda Z_2 \widetilde{\mathcal{I}}^{\rm (S)}_\lambda\left(\frac{m_1}{m} k_\gamma\right)\right],
\end{equation}
where
\begin{equation}\label{IES}
\widetilde{\mathcal{I}}^{\rm (S)}_\lambda (k)=\int^\infty_0 u^f_{l_f}u^i_{l_i}\varphi_\lambda(k\rho,\epsilon) d\rho
\end{equation}
and in the non-Siegert approach by
\begin{equation}\label{MEapp}
\langle \psi^f_{l_f}||\widetilde{\mathcal{M}}^{E}_{\lambda\mu}|| \psi^i_{l_i}\rangle=
 \frac{(2\lambda+1)!!}{k^{\lambda+1}_\gamma}\frac{ e\hbar}{2 \mu_M c}\mathbb{Y}_{l_f\lambda l_i} C^*_{l_f}C_{l_i}
\left[Z_1 \widetilde{\mathcal{I}}_\lambda \left(\frac{m_2}{m} k_\gamma\right)+(-1)^\lambda Z_2 \widetilde{\mathcal{I}}_\lambda\left(\frac{m_1}{m} k_\gamma\right)\right],
\end{equation}
where
\begin{equation}\label{IEapp}
\widetilde{\mathcal{I}}_\lambda (k)=\int^\infty_0 \frac{\partial \varphi_\lambda(k\rho,\epsilon) }{\partial\rho}
W_{fi}d\rho+[l_i(l_i+1)-l_f(l_f+1)]\int^\infty_0\frac{u^f_{l_f}u^i_{l_i}}{\rho^2}\varphi_\lambda(k\rho,\epsilon) d\rho
\end{equation}
and $W_{fi}$ designates the Wronskian of $u^f_{l_f}$ and $u^{i}_{l_i}$,
\begin{equation}\label{Wfi}
W_{fi}=u^f_{l_f} u^{i\prime}_{l_i}-u^{f\prime}_{l_f} u^i_{l_i}.
\end{equation}
For continuum to continuum transitions, the integrals in \Eqs{IES} and \eqref{IEapp} converge slowly.
Again, the contour integration method can be used for accelerating the convergence.

The reduced matrix elements of the Siegert multipole operators $\mathcal{M}^{E{\rm (S)}}_{\lambda\mu}$ between partial waves are given by
\begin{equation}\label{MESex}
\langle \psi^f_{l_f}||\mathcal{M}^{E{\rm (S)}}_{\lambda}|| \psi^i_{l_i}\rangle=\langle \psi^f_{l_f}||\widetilde{\mathcal{M}}^{E{\rm (S)}}_{\lambda}|| \psi^i_{l_i}\rangle
+\langle \psi^f_{l_f}||\mathcal{M}^{E}_{\lambda}|| \psi^i_{l_i}\rangle
-\langle \psi^f_{l_f}||\widetilde{\mathcal{M}}^{E}_{\lambda\mu}|| \psi^i_{l_i}\rangle.
\end{equation}
For real potentials, if the exact radial wave functions are considered, \Eqs{MES} and \eqref{MEapp} are equivalent and consequently, \Eqs{ME} and \eqref{MESex} are equivalent, too. 

The reduced matrix elements of the LWA Siegert multipole operator $\mathcal{M}^{E{\rm (S,LWA)}}_{\lambda}$ between partial waves are given by 
\begin{equation}\label{MElw}
\langle \psi^f_{l_f}||\mathcal{M}^{E{\rm (S,LWA)}}_{\lambda}|| \psi^i_{l_i}\rangle=
e Z^{(\lambda)}_{\rm eff}\mathbb{Y}_{l_f\lambda l_i} C^*_{l_f}C_{l_i}
\int^\infty_0 \rho^\lambda u^f_{l_f}u^i_{l_i} d\rho.
\end{equation}
For continuum to continuum transitions, the integrands behave asymptotically as an oscillating function times $\rho^\lambda$ with $\lambda>0$ and the integrals diverge, as anticipated in \Sec{Seccross}. 
To obtain a finite value, the technique used in~\Refs{TM85,LR86c,GJF12} is to replace the divergent integral by a limit of convergent integrals
\begin{equation}\label{MElwreg}
\langle \psi^f_{l_f}||\mathcal{M}^{E{\rm (S,LWA)}}_{\lambda}|| \psi^i_{l_i}\rangle_{\rm reg}=
e Z^{(\lambda)}_{\rm eff}\mathbb{Y}_{l_f\lambda l_i} C^*_{l_f}C_{l_i}
\lim\limits_{\epsilon \to 0} J_\lambda(\epsilon),
\end{equation}
where
\begin{equation}
J_\lambda(\epsilon)=\int^\infty_0 \rho^\lambda u^f_{l_f}u^i_{l_i} f(\epsilon,\rho) d\rho.
\end{equation}
The index reg is added to denote the regularized reduced matrix elements.
The regularization factor $f(\epsilon,\rho)$ is defined such that $J_\lambda(\epsilon)$ is finite for any strictly positive value of $\epsilon$, the limit of $J_\lambda(\epsilon)$ for $\epsilon\rightarrow0$ is finite, and
\begin{equation}
\lim\limits_{\epsilon \to 0} f(\epsilon,\rho)=1.
\end{equation}
More explicitly, the regularization factor is chosen to be an exponential in~\Refs{TM85,LR86c}, 
\begin{equation}\label{fchoixexp}
f(\epsilon,\rho)=e^{-\epsilon\rho}
\end{equation}
and a Gaussian in~\Ref{GJF12},
\begin{equation}\label{fchoixgauss}
f(\epsilon,\rho)=e^{-\epsilon^2\rho^2}.
\end{equation}
In the next section, it is proved that both choices of $f$ defined by \Eqs{fchoixexp} and \eqref{fchoixgauss} are equivalent.
The ways used in~\Refs{TM85,LR86c,GJF12} to evaluate the limit introduced in \Eq{MElwreg} are also explained. 
A new way to evaluate this limit, based on the contour integration method is also presented.

To conclude this section, let me note that the regularized reduced matrix elements defined by \Eq{MElwreg} can also be deduced from \Eqs{MES} and \eqref{IES} without introducing divergent integrals. 
Let me consider only the exponential and Gaussian choices of $\varphi_\lambda$, for which $\epsilon$ has a meaning.
At low photon energies, the reduced matrix elements of $\widetilde{\mathcal{M}}^{E{\rm (S)}}_{\lambda\mu}$ should be good approximations of the reduced matrix elements of $\mathcal{M}^{E{\rm (S)}}_{\lambda\mu}$ for values of $\epsilon$ small enough.
Since any value of $\epsilon$ which is strictly positive leads to convergent integrals $\mathcal{I}^{\rm (S)}_\lambda$, an arbitrary small value of 
$\epsilon$ can be considered, which is equivalent to take the limit for $\epsilon\rightarrow 0$ of $\mathcal{I}^{\rm (S)}_\lambda$.
For $\epsilon\rightarrow 0$, the reduced matrix element $\langle \psi^f_{l_f}||\widetilde{\mathcal{M}}^{E{\rm (S)}}_{\lambda}|| \psi^i_{l_i}\rangle$ tends to $\langle \psi^f_{l_f}||\mathcal{M}^{E{\rm (S,LWA)}}_{\lambda}|| \psi^i_{l_i}\rangle_{\rm reg}$, which justifies \Eq{MElwreg} without using the divergent expression \eqref{MElw}.
\section{Implementation of the regularization techniques}\label{Secreg}
\subsection{Fixed $\epsilon_0$ method}
The idea that Garrido, Jensen, and Fedorov have proposed in~\Ref{GJF12} is simply to approximate the limit for $\epsilon\rightarrow0$ by considering a small but finite value of $\epsilon$, denoted here by $\epsilon_0$,
\begin{equation}\label{ee}
\lim\limits_{\epsilon \to 0} J_\lambda(\epsilon)\approx J_\lambda(\epsilon_0).
\end{equation}
This approach is undeniably the simplest one.
It is applied easily for each multipole and does not require the calculation of the derivative of the radial wave functions.
Nevertheless, it appears to be unsatisfactory because, as noted in~\Ref{GJF12}, the value of $J_\lambda(\epsilon_0)$ is very sensitive to the value of $\epsilon_0$. 
To be acceptable, the choice of $\epsilon_0$ has to be such that any smaller value of $\epsilon_0$ leads to the same results, within the desired limits of accuracy.
For low photon energies, this criterion leads to very small values of $\epsilon_0$. 
However, the more $\epsilon_0$ is small, the more the integral $J_\lambda(\epsilon_0)$ converges slowly, which makes tedious its numerical integration.
In practice, to avoid a too slow convergence, the authors of~\Ref{GJF12} choose a rather big value of $\epsilon_0$, for which approximation \eqref{ee} can be very poor, as shown in~\Ref{GJF12} and in \Sec{Secres}.
Then, the bremsstrahlung cross sections $d\sigma/dE_\gamma$ are corrected by some more or less arbitrary cut and integrated to obtain the total bremsstrahlung cross section.
The total bremsstrahlung cross section seems stable with respect to small variations of $\epsilon_0$~\cite{GJF12} although the differential bremsstrahlung cross section cannot be considered as reliable.
The main drawback of this method is not to allow to obtain any reliable accurate differential bremsstrahlung cross sections, due to the fact that the values of $\epsilon_0$ considered in practice are chosen too big.
Both alternative methods presented in the next subsections do not have this inconvenience because they enable one to consider explicitly the case $\epsilon_0=0$.
\subsection{Integration by parts (IP) method}\label{TM}
This section presents a variant of the regularization technique proposed by Tanimura and Mosel and applied by them to the $E2$ operator in~\Ref{TM85}.
This variant has the advantage to be more easily generalizable to electric multipoles of any order.
Moreover, it can be applied for any potential $U$ singular at the origin, contrary to the version of Tanimura and Mosel.

The principle of the method is to derive, by some integration by parts and by using the properties of the radial wave functions, an expression of the function $J_\lambda$ which is valid and continuous at $\epsilon=0$.
Then, the limit for $\epsilon\rightarrow0$ is simply calculated by putting $\epsilon$ at zero in this expression.

Let me start by dividing the integral $J_\lambda$ into two integrals: from zero to $R$ ($R>0$) and from $R$ to infinity to avoid a particular treatment of potential $U$ singular at the origin,
\begin{equation}
J_\lambda(\epsilon)=
\int^R_0 \rho^\lambda u^f_{l_f}u^i_{l_i} f(\epsilon,\rho)  d\rho+
\int^\infty_R \rho^\lambda u^f_{l_f}u^i_{l_i} f(\epsilon,\rho) d\rho.
\end{equation}
The regularization method is based on the following relation
\begin{equation}\label{Eq1}
\begin{split}
E_\gamma \lim\limits_{\epsilon \to 0}  \int^{\infty}_R   g u^f_{l_f} u^i_{l_i}  f(\epsilon,\rho) d\rho=&
\lim\limits_{\epsilon \to 0}\left[
\beta \int^{\infty}_R g' W_{fi} f(\epsilon,\rho) d\rho+\int^{\infty}_R L g u^i_{l_i} u^f_{l_f} f(\epsilon,\rho) d\rho\right]\\
&+ \beta g(R) W_{fi}(R),
\end{split}
\end{equation}
where
\begin{equation}
L(\rho) = V_{l_i}(\rho)-V_{l_f}(\rho)
\end{equation}
and $g(\rho)$ is a function of class $C^{\infty}$ over $(R,\infty)$ with the asymptotic behavior $\rho^d$.
Roughly speaking, \Eq{Eq1} reduces the power of the divergence of the integral at $\epsilon=0$. 
Indeed, the integrand in the l.h.s.\ of \Eq{Eq1} behaves asymptotically as an oscillating function times $\rho^d$ whereas the integrands in the r.h.s.\ of \Eq{Eq1} behave asymptotically as oscillating functions times $\rho^{d-1}$ (for $d\neq 0$) and $\rho^{d-2}$, respectively. 
The case $d=0$ is even more favorable.

The power of the divergence of the second integral of the r.h.s\ of \Eq{Eq1} can be reduced by applying recursively \Eq{Eq1}. 
For reducing the power of the divergence of the first integral of the r.h.s\ of \Eq{Eq1}, the following relation can be used
\begin{equation}\label{Eq2}
\begin{split}
 E_\gamma \lim\limits_{\epsilon \to 0}\int^{\infty}_R g W_{fi} f(\epsilon,\rho) d\rho = 
\lim\limits_{\epsilon \to 0} \left\{ \int^{\infty}_R  \left[ g h'+2g' h-\beta g''' \right]u^f_{l_f} u^i_{l_i}  f(\epsilon,\rho) d\rho\right. \\
\left.+\int^{\infty}_R L g W_{fi} f(\epsilon,\rho) d\rho\right\}
+\int^{\infty}_R  ( 2 g U'+4 g' U)u^f_{l_f} u^i_{l_i}  d\rho+X_g(R),
\end{split}
\end{equation}
where
\begin{eqnarray}\label{defh}
h(\rho) &=& V_{l_i}(\rho)+V_{l_f}(\rho)+2\frac{Z_1 Z_2 e^2}{\rho}-E_i-E_f,\\
X_g(\rho) &=& \left[g (h+2U)-\beta g''\right] u^i_{l_i}u^f_{l_f}-2\beta g u^{i\prime}_{l_i}u^{f\prime}_{l_f}+\beta g'(u^f_{l_f} u^{i}_{l_i})'.
\end{eqnarray}

Like \Eq{Eq1}, \Eq{Eq2} reduces the power of the divergence of the integral at $\epsilon=0$.
The integrand in the l.h.s.\ of \Eq{Eq1} behaves asymptotically as an oscillating function times $\rho^d$ whereas the integrands of the first two integrals in the r.h.s.\ of \Eq{Eq1} behave asymptotically as oscillating functions times $\rho^{d-1}$ (for $d\neq 0$) and $\rho^{d-2}$, respectively. The case $d=0$ is more favorable, again.
The last integral converges without the regularization factor since the potential $U$ has a finite range.
\Eqs{Eq1} and \eqref{Eq2} are inspired from equations (A.5) and (A.6) in~\Ref{TM85}. 
They are proved in the Appendix.

By applying recursively \Eqs{Eq1} and \eqref{Eq2} to $J_\lambda$, one obtains an expression which converges at $\epsilon=0$ having the following form
\begin{equation}\label{EqTM}
\lim\limits_{\epsilon \to 0} J_\lambda(\epsilon)=\int^R_0 \rho^\lambda u^f_{l_f}u^i_{l_i} d\rho+ \int^\infty_R C^{(\lambda)}_1 u^f_{l_f} u^i_{l_i}  d\rho
+\int^\infty_R C^{(\lambda)}_2 W_{fi}  d\rho+C^{(\lambda)}_3(R).
\end{equation}
Since \Eqs{Eq1} and \eqref{Eq2} are valid for both choices of $f(\epsilon,\rho)$, defined by \Eqs{fchoixexp} and \eqref{fchoixgauss}, \Eq{EqTM} is valid for both choices of the regularization function (exponential or Gaussian).

The coefficients $C^{(\lambda)}_1$, $C^{(\lambda)}_2$, and $C^{(\lambda)}_3$ are given explicitly for $E1$ by
\begin{eqnarray}\label{C11}
C^{(1)}_1 (\rho)&=& \frac{1}{E_\gamma}\left(L\rho+ \beta \frac{h'+2U'}{E_\gamma}\right),\\
C^{(1)}_2 (\rho)&=&  \frac{\beta L}{E^2_\gamma}, \\
C^{(1)}_3 (\rho) &=& \frac{\beta}{E_\gamma} \left(\rho W_{fi}+\frac{X_1}{E_\gamma}\right)
\end{eqnarray}
and for $E2$ by
\begin{eqnarray}
C^{(2)}_1 (\rho)&=& \frac{1}{E^2_\gamma}\left[2\beta(h' \rho +2 U' \rho+4U)+L^2\rho^2+4\beta\frac{Lh}{E_\gamma}\right],\\
C^{(2)}_2 (\rho)&=& \frac{2\beta}{E^2_\gamma}\left(L\rho+2\beta \frac{h'}{E_\gamma}
\right),\\ \label{C23}
C^{(2)}_3 (\rho)&=& \frac{\beta}{E_\gamma}W_{fi}\left(
\rho^2+ \frac{L\rho^2}{E_\gamma}+4 \beta\frac{h}{E^2_\gamma}\right)+2\beta\frac{X_\rho}{E^2_\gamma}.
\end{eqnarray}
The expressions \eqref{C11}-\eqref{C23} are derived by applying recursively the transformations defined by \Eqs{Eq1} and \eqref{Eq2} until convergent integrals are obtained at $\epsilon=0$.
Extra transformations might be performed for improving the convergence rate of the integrals. 
In this case, the coefficients $C^{(\lambda)}_1$, $C^{(\lambda)}_2$, and $C^{(\lambda)}_3$ are more complex but lead to the same results except for the numerical accuracy.
If the potential $U$ is non singular at the origin, $R$ can be taken as zero. 
In this case, the terms $C^{(1)}_3$ and $C^{(2)}_3$ are null.

A comparison between \Eqs{IE} and \eqref{EqTM} shows that the IP approach is not numerically more advantageous, on the contrary, than the non-Siegert approach.
Both methods require the calculation of the derivative of the radial wave functions but the integrands in the IP approach are more complicated than in the non-Siegert approach.
Moreover, this complexity increases with the order of the multipole in the IP approach while it does not change in the non-Siegert approach.
\subsection{Contour integration (CI) method}\label{subSeccontour}
Like the IP method, the CI method aims at deriving an expression of the function $J_\lambda$ which is valid and continuous at $\epsilon=0$.
In the CI method, this expression is obtained from the contour integration method proposed in~\Ref{VF70}.

\begin{figure}[ht]
\begin{center}
\includegraphics[scale=0.8]{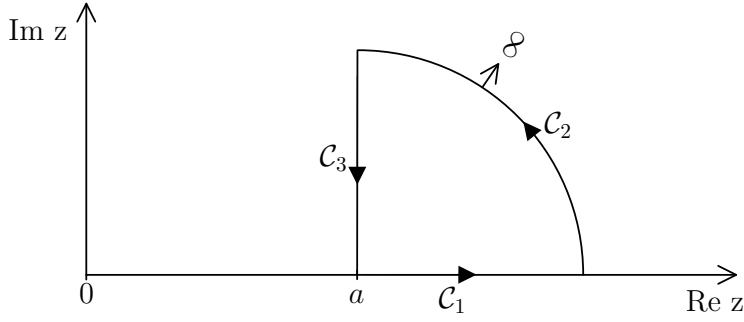}
\end{center}
\caption{Schematic view of the contour of integration in the complex $z$-plane.}  
\label{Figcont}
\end{figure}

Let me divide the integral $J_\lambda$ into two regions: an inner region ($\rho<a$) and an external region ($\rho>a$), where the radial wave functions can be replaced by their asymptotic form with a good accuracy,
\begin{equation}\label{sepJ}
J_\lambda(\epsilon)=
\int^a_0 \rho^\lambda u^f_{l_f}u^i_{l_i} f(\epsilon,\rho)  d\rho+
\int^\infty_a \rho^\lambda u^{f,{\rm as}}_{l_f}u^{i,{\rm as}}_{l_i} f(\epsilon,\rho) d\rho.
\end{equation}
Contrary to parameter $R$, which divides the space integration in the IP approach, parameter $a$ cannot be arbitrary small.
It has to be large enough for that the effects of the potential $U$ can be neglected in the external region.
From \Eq{as2}, the second integral can be written as
\begin{equation}\label{sep}
\int^\infty_a \rho^\lambda u^{f,{\rm as}}_{l_f}u^{i,{\rm as}}_{l_i} f(\epsilon,\rho) d\rho=\frac{(1+\delta_{12})^{1/2}}{\sqrt{v_i} k_i} 
{\rm Im}\left[e^{i\delta_{l_i}}\int^\infty_a  \rho^\lambda  u^{f,{\rm as}}_{l_f}(\rho) O_{l_i}(\eta_i,k_i\rho)  f(\epsilon,\rho) d\rho\right],
\end{equation}
where ${\rm Im}[...]$ is the imaginary part of the complex number between brackets.
First, let me consider the exponential regularization factor, defined by \Eq{fchoixexp}, because it leads to a simpler expression.
An expression of $J_\lambda(\epsilon)$ valid for exponential and Gaussian regularization factors is derived farther.
In the case where $f$ is an exponential, the r.h.s.\ integral in \Eq{sep} can be evaluated from the following contour integral in the complex $z$-plane,
\begin{equation} \label{intcont}
\int_\mathcal{C} z^\lambda  u^{f,{\rm as}}_{l_f}(z) O_{l_i}(\eta_i,k_i z)  e^{-\epsilon z} dz.
\end{equation}
The contour $\mathcal{C}$ is schematically represented in \Fig{Figcont}, where it is divided into three parts: $\mathcal{C}=\mathcal{C}_1+\mathcal{C}_2+\mathcal{C}_3$.
It is explicitly defined by
\begin{align}
 z=
 \begin{cases}
  x &{\rm with} \ a\leq x<\infty\ {\rm over}\ \mathcal{C}_1,\\
  a+\mathcal{R} e^{i\varphi} &{\rm with}\ \mathcal{R}\rightarrow\infty\ {\rm and}\ 0< \varphi<\pi/2\ {\rm over}\ \mathcal{C}_2,\\
    a+iy  &{\rm with} \ 0< y<\infty\ {\rm over}\ \mathcal{C}_3.\\
  \end{cases}
\end{align}

Since the integrand is regular inside the region delimited by contour $\mathcal{C}$, the integral over $\mathcal{C}$ is null. 
The dominant part of the oscillating terms in integral \eqref{intcont} behaves asymptotically as $e^{i(k_i\pm k_f)\rho}$. 
Since $k_i$ is larger than $k_f$, $k_i\pm k_f$ is positive and the integral over $\mathcal{C}_2$ is null.
The integral over $\mathcal{C}_1$ is thus equal to the opposite of the integral over $\mathcal{C}_3$,
\begin{equation}\label{contour}
\int^\infty_a  \rho^\lambda  u^{f,{\rm as}}_{l_f}(\rho) O_{l_i}(\eta_i,k_i\rho) e^{-\epsilon \rho} d\rho=
i\int^\infty_0  z_+^\lambda  u^{f,{\rm as}}_{l_f}(z_+) O_{l_i}(\eta_i,k_i z_+) e^{-\epsilon z_+} dy
\end{equation}
with $z_+=a+iy$.
The transformation defined by \Eq{contour} replaces the imaginary exponentials of the initial integrand by decreasing exponentials.
From \Eqs{sepJ}, \eqref{sep}, and \eqref{contour}, $J_\lambda(\epsilon)$ can be written for any strictly positive value of $\epsilon$ as
\begin{equation}\label{Jexp}
J_{\lambda{,\rm exp}}(\epsilon)=\int^a_0 \rho^\lambda u^f_{l_f}u^i_{l_i} e^{-\epsilon\rho}  d\rho+\frac{(1+\delta_{12})^{1/2}}{\sqrt{v_i} k_i} 
{\rm Im}\left[i e^{i\delta_{l_i}}\int^\infty_0  z_+^\lambda  u^{f,{\rm as}}_{l_f}(z_+) O_{l_i}(\eta_i,k_i z_+)  e^{-\epsilon z_+} dy\right].
\end{equation}
The subscript exp is added to $J_\lambda$ to recall that this expression is only valid when $f$ is an exponential.
In the r.h.s.\ of \Eq{Jexp}, the exponentials $e^{-\epsilon\rho}$ and $e^{-\epsilon z_+}$ are not required to ensure the convergence at $\epsilon=0$.
The r.h.s.\ of \Eq{Jexp} defines a function continuous at $\epsilon=0$.
The limit for $\epsilon=0$ is thus evaluated straightforwardly,
\begin{equation}\label{Jreg}
\lim\limits_{\epsilon \to 0}J_\lambda(\epsilon)=\int^a_0 \rho^\lambda u^f_{l_f}u^i_{l_i}   d\rho+\frac{(1+\delta_{12})^{1/2}}{\sqrt{v_i} k_i} 
{\rm Im}\left[i e^{i\delta_{l_i}} \int^\infty_0  z_+^\lambda  u^{f,{\rm as}}_{l_f}(z_+) O_{l_i}(\eta_i,k_i z_+) dy\right].
\end{equation}
The subscript exp is here dropped because this equation is also valid for a Gaussian choice of $f$ contrary to \Eq{Jexp}.
Indeed, a proof is given in \Sec{TM} that the limit for $\epsilon\rightarrow 0$ is independent of the choice of $f$, exponential or Gaussian.
An alternative proof based on the contour integration method is developed hereafter.
Let me note that if the Gaussian regularization factor is considered in the contour integral \eqref{intcont}, the integrals over $\mathcal{C}_2$ and $\mathcal{C}_3$ are infinite.
To calculate the r.h.s.\ integral of \Eq{sep} by a contour integration method valid for both exponential and Gaussian regularization functions, the substitution $t=\rho^2$ is made
\begin{equation}\label{esub}
\int^\infty_a  \rho^\lambda  u^{f,{\rm as}}_{l_f}(\rho) O_{l_i}(\eta_i,k_i\rho) f(\epsilon,\rho) d\rho=\frac{1}{2}
\int^\infty_{a^2}  t^{(\lambda-1)/2}  u^{f,{\rm as}}_{l_f}(\sqrt{t}) O_{l_i}(\eta_i,k_i\sqrt{t}) f(\epsilon,\sqrt{t}) dt.
\end{equation}
The r.h.s.\ integral can be evaluated from the integral over $\mathcal{C}'$
\begin{equation}
\int_{\mathcal{C}'}  z^{(\lambda-1)/2}  u^{f,{\rm as}}_{l_f}(\sqrt{z}) O_{l_i}(\eta_i,k_i\sqrt{z})  f(\epsilon,\sqrt{z}) dz,
\end{equation}
where the contour $\mathcal{C}'$ is defined in the same way as $\mathcal{C}$ except that $a$ is replaced by $a^2$. 
By convention, $\sqrt{z}$ is the complex number such that its square is $z$ and its real part is positive and $z^{(\lambda-1)/2}=(\sqrt{z})^{\lambda-1}$. 
Since the integrals over $\mathcal{C}'$ and $\mathcal{C}'_2$, obtained from $\mathcal{C}_2$ by replacing $a$ by $a^2$, are null, one has
\begin{equation}\label{eqG}
\int^\infty_{a^2} t^{(\lambda-1)/2}  u^{f,{\rm as}}_{l_f}(\sqrt{t}) O_{l_i}(\eta_i,k_i\sqrt{t}) f(\epsilon,\sqrt{t}) dt=
i\int^\infty_0 \tilde{z}^{(\lambda-1)/2}  u^{f,{\rm as}}_{l_f}(\sqrt{\tilde{z}}) O_{l_i}(\eta_i,k_i\sqrt{\tilde{z}}) f(\epsilon,\sqrt{\tilde{z}}) dy
\end{equation}
with $\tilde{z}=a^2+i y$. 
From \Eqs{sepJ}, \eqref{sep}, \eqref{esub}, and \eqref{eqG}, $J_\lambda(\epsilon)$ can be written for any strictly positive value of $\epsilon$ as
\begin{equation}\label{JG}
\begin{split}
J_\lambda(\epsilon)=&\int^a_0 \rho^\lambda u^f_{l_f}u^i_{l_i}f(\epsilon,\rho) d\rho\\
&+\frac{(1+\delta_{12})^{1/2}}{2\sqrt{v_i} k_i} 
{\rm Im}\left[i e^{i\delta_{l_i}}
\int^\infty_0 \tilde{z}^{(\lambda-1)/2}  u^{f,{\rm as}}_{l_f}(\sqrt{\tilde{z}}) O_{l_i}(\eta_i,k_i\sqrt{\tilde{z}}) f(\epsilon,\sqrt{\tilde{z}})  dy
\right].
\end{split}
\end{equation}
The r.h.s.\ integral in \Eq{JG} converges even without the regularization factor and one has
\begin{equation}\label{ee2}
\lim\limits_{\epsilon \to 0} J_\lambda(\epsilon)=
\int^a_0 \rho^\lambda u^f_{l_f}u^i_{l_i} d\rho+
\frac{(1+\delta_{12})^{1/2}}{2\sqrt{v_i} k_i} 
{\rm Im}\left[i e^{i\delta_{l_i}}
\int^\infty_0 \tilde{z}^{(\lambda-1)/2}  u^{f,{\rm as}}_{l_f}(\sqrt{\tilde{z}}) O_{l_i}(\eta_i,k_i\sqrt{\tilde{z}})dy
\right].
\end{equation}
Since \Eq{ee2} is valid for both choices of $f$, defined by \Eqs{fchoixexp} and \eqref{fchoixgauss}, it proves in an alternative way that the results should not depend on the particular choice of the regularization function, exponential or Gaussian.

\section{$\alpha+\alpha$ bremsstrahlung}\label{Secres}
The potential models are applied to the $\alpha+\alpha$ bremsstrahlung. 
Since the $\alpha$ particles are bosons, odd-parity multipoles are forbidden. 
Moreover, M1 transitions are also forbidden at the long-wavelength approximation because of the orthogonality 
between the initial and final states~\cite{BD83}.
Only the E2 transitions, which are dominant, are considered here. 

Many forms of the electric transition multipole operators are presented in \Sec{Secelec}. 
However, it is not required to consider all of them because, as explained in \Sec{Secelec}, some are equivalent.
In practice, only the models based on the non-Siegert operator $\mathcal{M}^{E}_{\lambda\mu}$, the approximate Siegert operator $\widetilde{\mathcal{M}}^{E{\rm (S)}}_{\lambda\mu}$ corresponding to the Bessel choice, and the regularized expressions of the LWA operator $\widetilde{\mathcal{M}}^{E{\rm (S,LWA)}}_{\lambda\mu}$, calculated with the CI method, are studied.

The interaction between the $\alpha$ nuclei is described by the BFW potential~\cite{BFW77}, like in several previous calculations of the $\alpha+\alpha$ bremsstrahlung \cite{LR86,La86,LR86c,BSD91}.
The BFW potential reproduces the experimental $S$, $P$, and $D$ phase shifts up to $20$~MeV.
This is the sum of a deep Gaussian and a screened Coulomb potential, 
\begin{equation}
V(\rho)=-V_0 e^{-\nu \rho^2}+\frac{4 e^2}{\rho} \erf(\xi \rho),
\end{equation}
where $\erf$ is the error function.
The parameters $\nu$ and $\xi$ are set at $\nu=0.22$~fm$^{-2}$, and $\xi=0.75$~fm$^{-1}$ as in \Ref{BFW77}.
The parameter $V_0$ is set at $V_0=122.61$~MeV. 
With these values and $\beta=10.368~$MeV~fm$^2$, the exerimental resonance at $92$~keV in the $0^+$ phase shift is reproduced by the potential model with a precision of $1$~keV.

The radial wave functions are obtained by solving \Eq{Eqrad} with a Numerov algorithm~\cite{Ra72}. 
All radial integrals required to calculate the reduced matrix elements of the electric transition multipole operators are evaluated by the contour integration approach.
The integrals over the real axis, from $0$ to $a$, are evaluated with the Weddle's rule \cite{WR24} while the integrals over the imaginary axis are evaluated by a Gauss-Laguerre quadrature associated with a suitable scale factor. 
The Coulomb functions are calculated, over the real axis, by the routine described in \Ref{BFS74} and, over the imaginary axis, by their asymptotic expansions \cite{AS65} or by the routine described in \Ref{TB85}.

The multipole matrix elements $u^{\rm E}_{\lambda\mu}$ is evaluated from its partial-wave expansion, given by \Eq{upart}, truncated at $l_i=l_f=l_{max}$. 
For low-photon energies, series \eqref{upart} converges (very) slowly and a large value of $l_{max}$ is required to reach convergence~\cite{BSD91}.
As explained in~\Ref{BSD91}, the convergence of this series can be accelerated by a Kummer's series transformation~\cite{AS65}.
However, this convergence acceleration method is currently applicable only to the non-Siegert approach. 
By consistency, this method is not applied in this work and the three approaches are compared for the same values of $l_{max}$.
In all cases, it is verified that adding some extra partial waves beyond $l_{max}$ in the evaluation of $u^{\rm E}_{\lambda\mu}$ implies a relative modification of the bremsstrahlung cross sections $d\sigma/dE_\gamma$ smaller than $1\%$. 

\begin{figure}[!ht]
\centering
\includegraphics[scale=0.7]{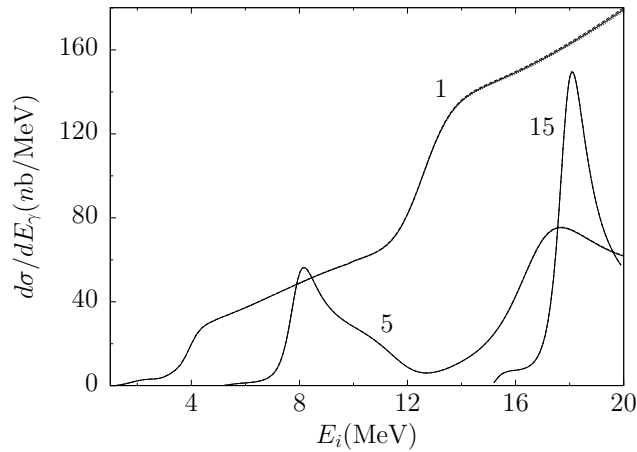}
\caption{Bremsstrahlung cross sections $d\sigma/dE_\gamma$ corresponding to the operator $\mathcal{M}^{E}_{\lambda\mu}$ (full lines), 
to the operator $\widetilde{\mathcal{M}}^{E{\rm (S)}}_{\lambda\mu}$ (dashed lines), and to the operator $\widetilde{\mathcal{M}}^{E{\rm (S,LWA)}}_{\lambda\mu}$ (dots) as a function of the initial energy $E_i$ for three photon energies: $E_\gamma=1$, $5$, and $15$~MeV.
Differences between full lines, dashed lines, and dots are nearly indistinguishable at the scale of the figure.}
\label{fig2}
\end{figure}

In \Fig{fig2}, bremsstrahlung cross sections $d\sigma/dE_\gamma$ obtained with the non-Siegert model ($\mathcal{M}^{E}_{\lambda\mu}$), the approximate Siegert model ($\widetilde{\mathcal{M}}^{E{\rm (S)}}_{\lambda\mu}$) and the LWA model ($\widetilde{\mathcal{M}}^{E{\rm (S,LWA)}}_{\lambda\mu}$) are compared for an initial energy $E_i$ up to $20$~MeV and three values of the photon energy: $E_\gamma=1$, $5$, and $15$~MeV.
For $E_{\gamma}=5$~MeV and $E_{\gamma}=15~$MeV, series in \Eq{upart} is truncated at $l_{max}=10$ and $l_{max}=6$, respectively.
For $E_{\gamma}=1$~MeV, the value of $l_{max}$ depends on the initial energy of the collision. 
It varies from $10$ at $E_i=1.2$~MeV to $60$ at $E_i=20$~MeV.
\Fig{fig2} shows that the three approaches (non-Siegert, approximate Siegert, and LWA) lead to nearly identical bremsstrahlung cross sections for large ranges of colliding energy and different photon energies.

\begin{figure}[!ht]
\centering
\includegraphics[scale=0.7]{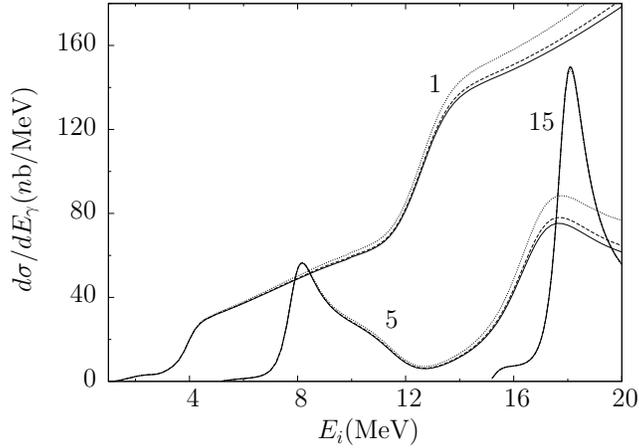}
\caption{Bremsstrahlung cross sections $d\sigma/dE_\gamma$ at $E_\gamma=1$, $5$, and $15$~MeV at the long-wavelength approximation. 
Full lines correspond to $\epsilon\rightarrow0$, dashed lines correspond to $\epsilon_0=0.001$~fm$^{-1}$ for $E_\gamma=1$~MeV and to $\epsilon_0=0.01$~fm$^{-1}$ for $E_\gamma=5$ and $15$~MeV, and dots correspond to $\epsilon_0=0.002$~fm$^{-1}$ for $E_\gamma=1$~MeV and to $\epsilon_0=0.02$~fm$^{-1}$ for $E_\gamma=5$ and $15$~MeV. The Gaussian regularization function is considered.}
\label{fig3}
\end{figure}

For the same energies, the bremsstrahlung cross sections $d\sigma/dE_\gamma$ obtained in the LWA approach by using a non-zero value of $\epsilon$, i.e.\ by evaluating reduced matrix elements \eqref{MElwreg} with the fixed $\epsilon_0$ model [\Eq{ee}], are shown in \Fig{fig3}.
They are compared with the LWA bremsstrahlung cross sections obtained for $\epsilon\rightarrow 0$. 
The Gaussian regularization function is considered here. 
The case of an exponential regularization function is briefly discussed farther.

The values of $\epsilon_0$ used in \Refs{GJF12,GJF14} to calculate the integrated bremsstrahlung cross sections are $\epsilon_0=0.02$~fm$^{-1}$ \cite{GJF12} and $\epsilon_0=0.01$~fm$^{-1}$ \cite{GJF14}.
The same values are considered for $E_\gamma=5$ and $15$~MeV in \Fig{fig3}.
At $E_\gamma=15$~MeV, both considered values of $\epsilon_0$ give accurate results. 
Differences between bremsstrahlung cross sections for $\epsilon\rightarrow 0$ and the fixed $\epsilon_0$ method are around $0.4$~nb/MeV for $\epsilon_0=0.01$~fm$^{-1}$ and $1.6$~nb/MeV for $\epsilon_0=0.02$~fm$^{-1}$ at the peak ($E_i\approx 18$~MeV). 
They are (much) smaller at the other considering colliding energies.
At $E_\gamma=5$~MeV, the bremsstrahlung cross sections for $\epsilon\rightarrow 0$ are less well approximated by bremsstrahlung cross sections obtained by using $\epsilon_0=0.01$ and $0.02$~fm$^{-1}$ than at $E_\gamma=15$~MeV. 
The gap between bremsstrahlung cross sections obtained when $\epsilon\rightarrow 0$ and the ones obtained by using a fixed value of $\epsilon_0$ goes up to around $3$~nb/MeV for $\epsilon_0=0.01$~fm$^{-1}$ and around $15$~nb/MeV for $\epsilon_0=0.02$~fm$^{-1}$ at $E_i=20$~MeV.
At $E_\gamma=1$~MeV, the bremsstrahlung cross sections obtained with $\epsilon_0=0.01$ and $0.02$~fm$^{-1}$ (not shown) are in full disagreement with the bremsstrahlung cross sections obtained when $\epsilon\rightarrow 0$. 
Moreover, contributions of the partial waves beyond $l_{max}=60$, negligible in the (approximate) Siegert, non-Siegert, and LWA models, are surprisingly large, which has no physical meaning. 
This is the reason why, for $E_\gamma=1$~MeV, the bremsstrahlung cross sections $d\sigma/dE_\gamma$ obtained with values of $\epsilon_0$ ten times smaller ($\epsilon_0=0.001$ and $0.002$~fm$^{-1}$) are displayed in \Fig{fig3}.
Even for these much smaller values of $\epsilon_0$, it is noted that the gap with the bremsstrahlung cross sections obtained when $\epsilon\rightarrow 0$ is significant.
It is thus deduced from \Fig{fig3} that the smaller the photon energy is, the smaller $\epsilon_0$ has to be taken.

\begin{figure}[!ht]
\centering
\includegraphics[scale=0.7]{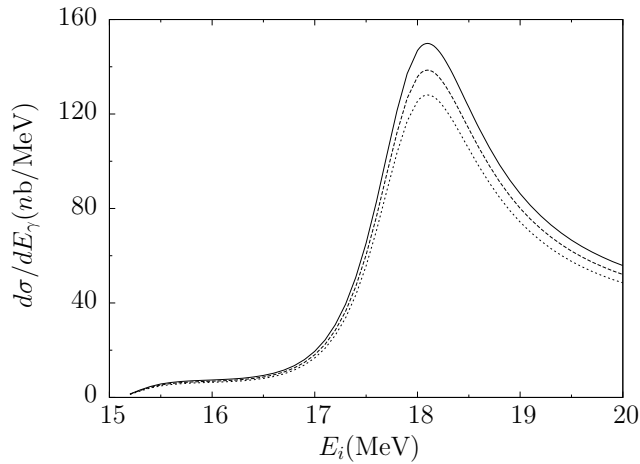}
\caption{Bremsstrahlung cross sections $d\sigma/dE_\gamma$ at $E_\gamma=15$~MeV at the long-wavelength approximation. 
Full lines correspond to $\epsilon\rightarrow0$, dashed lines correspond to $\epsilon_0=0.01$~fm$^{-1}$, and dots correspond to $\epsilon_0=0.02$~fm$^{-1}$. The exponential regularization function is considered.}
\label{fig4}
\end{figure}

A similar statement can be done by considering the exponential regularization function. 
However, for the same accuracy, smaller values of $\epsilon_0$ have to be considered as it can be seen in \Fig{fig4}, where the bremsstrahlung cross sections corresponding at $E_\gamma=15$~MeV obtained with $\epsilon_0=0.01$ and $0.02$~fm$^{-1}$ are displayed.

\begin{figure}[!ht]
\centering
\includegraphics[scale=0.7]{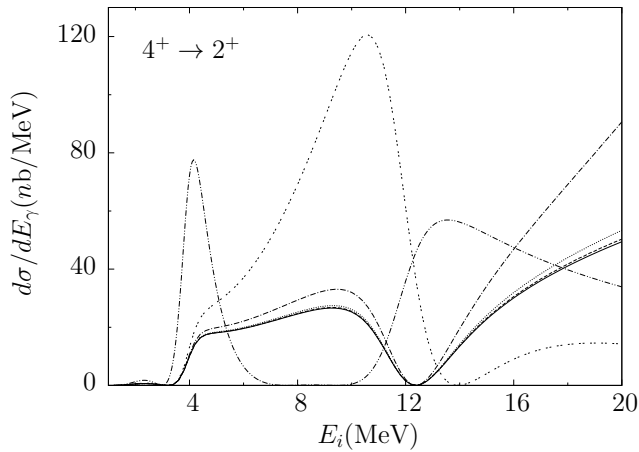}
\caption{Bremsstrahlung cross sections $d\sigma/dE_\gamma$ at $E_\gamma=1$~MeV at the long-wavelength approximation by considering only the $4^+\rightarrow 2^+$ transition. 
They are evaluated for $\epsilon\rightarrow0$ (full), $\epsilon_0=0.001$~fm$^{-1}$ (dash), $\epsilon_0=0.002$~fm$^{-1}$ (dot), $\epsilon_0=0.005$~fm$^{-1}$ (dash-dot), $\epsilon_0=0.01$~fm$^{-1}$ (dash-dash), and $\epsilon_0=0.02$~fm$^{-1}$ (dash-dot-dot). The Gaussian regularization function is considered.}
\label{fig5}
\end{figure}

To illustrate the effects of choosing bigger values of $\epsilon_0$ than $\epsilon_0=0.001$ and $0.002$~fm$^{-1}$ at $E_\gamma=1$~MeV without being concerned by the truncation problem of series \eqref{upart}, the bremsstrahlung cross sections can be evaluated for a specific single transition. 
In \Fig{fig5}, bremsstrahlung cross sections are evaluated by considering only the $4^+\rightarrow 2^+$ transition. 

Five values of $\epsilon_0$ ($0.001$, $0.002$, $0.005$, $0.01$, and $0.02$~fm$^{-1}$) are considered and the results are compared to the limit case $\epsilon\rightarrow 0$. At $E_\gamma=1$~MeV, the bremsstrahlung cross sections obtained with $\epsilon_0=0.01$ and $0.02$~fm$^{-1}$ can definitively not be considered as reliable approximations of the bremsstrahlung cross sections obtained at the limit for $\epsilon\rightarrow 0$.
\section{Conclusion}\label{Secconcl}
Different potential models of nucleus-nucleus bremsstrahlung are presented and compared. 
These models differ by the form of the electric transition multipole operators which are used. 
In the non-Siegert model, the electric transition multipole operators are written from the nuclear current density.
In the Siegert model, a part of the current dependence of the electric transition multipole operators, expected to be dominant at low photon energy, is replaced by a term depending on the charge density.
Considering only the charge-dependent term defines the approximate Siegert model.
Then, making the long-wavelength approximation defines the long-wavelength approximation (LWA) model.
Contrary to the other models, the LWA model leads to divergent integrals in the bremsstrahlung calculations and thus requires using some regularization techniques to obtain finite values for the bremsstrahlung cross sections.

The models are applied to the $\alpha+\alpha$ bremsstrahlung for a range of energies where the description of the $\alpha+\alpha$ scattering by a potential model is accurate. 
The $\alpha\alpha$ interaction is described by a real purely central potential. In this case, the Siegert and non-Siegert models are proved to be equivalent.
For the $\alpha+\alpha$ system and the considered energies, there is no significant difference between the bremsstrahlung cross sections obtained with the equivalent Siegert and non-Siegert models and the ones obtained with the approximate Siegert model.
The LWA model also leads to nearly identical results as long as a proper regularization technique is used.

Three regularization techniques are presented: the fixed $\epsilon_0$ method, the integration by parts (IP) method, and the contour integration (CI) method. 
The limits of validity of the fixed $\epsilon_0$ method are discussed theoretically and exemplified on the $\alpha+\alpha$ bremsstrahlung.
The IP method leads to quite complicated expressions which makes its use tedious in particular if high multipole orders are considered.
The CI method is proved to be a particularly convenient and efficient method to regularize the divergent bremsstrahlung matrix elements.
\appendix*
\section{Derivation of \Eqs{Eq1} and \eqref{Eq2}}
Let $G$ be a function of class $C^1$ over an interval $(R,b)$ with $b\geq R$.
One proves from \Eqs{consen} and \eqref{Eqrad} and by using an integration by parts that
\begin{eqnarray}
E_\gamma \int^b_R G u^f_{l_f} u^i_{l_i} d\rho &=& \int^b_R Gu^f_{l_f} E_i u^i_{l_i} d\rho-
 \int^b_R Gu^i_{l_i} E_f u^f_{l_f} d\rho \\
&=& -\beta \int^b_R G W^{\prime}_{fi} d\rho
+\int^b_R L G  u^i_{l_i} u^f_{l_f} d\rho\\ \label{Aeq1}
&=& -\beta[G W_{fi}]^b_R
+ \beta \int^b_R  G' W_{fi} d\rho+\int^b_R L G  u^i_{l_i} u^f_{l_f} d\rho,
\end{eqnarray}
where $W_{fi}$ is the Wronskian of $u^f_{l_f}$ and $u^{i}_{l_i}$ defined by \Eq{Wfi} and $\left[F\right]^b_R$ is a shorthand notation for $F(b)-F(R)$.

Let me note that by considering $G=\varphi_\lambda(k\rho,\epsilon)$ and $R=0$ and by taking the limit for $b\rightarrow\infty$, \Eq{Aeq1} proves the equivalence between the reduced matrix elements of $\widetilde{\mathcal{M}}^{E{\rm (S)}}_{\lambda}$ and $\widetilde{\mathcal{M}}^E_{\lambda}$ evaluated between partial waves, which are given by \Eqs{MES} and \eqref{MEapp}.

Now, let me choose $G=g f(\epsilon,\rho)$ where $\epsilon>0$ and $g$ is a function of class $C^1$ over $(R,\infty)$ and let me take the limit for $b\rightarrow\infty$ in \Eq{Aeq1}. 
Every term in \Eq{Aeq1} is assumed to have a finite limit for $b\rightarrow\infty$.
For the exponential regularization factor, the first integral of the r.h.s.\ can be written as 
\begin{equation}\label{eqa1}
\int^\infty_R  (g e^{-\epsilon\rho})' W_{fi}  d\rho=\int^\infty_R  g' W_{fi}e^{-\epsilon\rho} d\rho -\epsilon\int^\infty_R  g W_{fi}e^{-\epsilon\rho} d\rho.
\end{equation}
By using a contour integration like in~\Ref{GV71}, the limit for $\epsilon\rightarrow 0$ of the last integral is proved to be bounded and then,
\begin{equation}
\lim\limits_{\epsilon \to 0} \epsilon\int^\infty_R  g W_{fi}e^{-\epsilon\rho} d\rho=0.	
\end{equation}
A similar derivation can be done for the Gaussian regularization factor. 
One thus has for both regularization factors
\begin{equation}\label{eqa3}
\lim\limits_{\epsilon \to 0} \int^\infty_R  G' W_{fi} d\rho=\lim\limits_{\epsilon \to 0} \int^\infty_R  g' W_{fi}f(\epsilon,\rho) d\rho.
\end{equation}
By taking the limit for $\epsilon\rightarrow 0$ of both sides of \Eq{Aeq1} and by using \Eq{eqa3}, \Eq{Eq1} is obtained.

For proving \Eq{Eq2}, let me start by differentiating the radial Schr\"odinger equation \eqref{Eqrad} with respect to $\rho$
\begin{equation}\label{Eqradder}
-\beta u^{c\prime\prime\prime}_{l_c}(\rho)+ \left[V_{l_c}(\rho)+V(\rho)\right] u^{c\prime}_{l_c}(\rho) + \left[V^{\prime}_{l_c}(\rho)+V'(\rho)\right] u^{c}_{l_c}(\rho)=E_c u^{c\prime}_{l_c}(\rho).
\end{equation}
The function $G$ is assumed here to be a function of class $C^3$ over an interval $(R,b)$ with $b\geq R$.
From \Eqs{consen}, \eqref{Eqrad}, and \eqref{Eqradder}, one proves that
\begin{eqnarray}
E_\gamma \int^b_R G  W_{fi} d\rho &=&
\int^b_R G (u^f_{l_f} E_i u^{i\prime}_{l_i}- u^{i\prime}_{l_i} E_f u^f_{l_f}-u^{f\prime}_{l_f} E_i u^i_{l_i}+u^i_{l_i} E_f u^{f\prime}_{l_f}) d\rho\\\nonumber
&=& \beta \int^b_R G (u^{i\prime\prime}_{l_i}u^{f\prime}_{l_f}+u^{f\prime\prime}_{l_f}u^{i\prime}_{l_i}
-u^{i\prime\prime\prime}_{l_i}u^{f}_{l_f}-u^{f\prime\prime\prime}_{l_f} u^i_{l_i}) d\rho \\ \label{Aeq2}
&&+\int^b_R G u^f_{l_f} u^{i}_{l_i} (h'+2U') d\rho+\int^b_R L G W_{fi} d\rho,
\end{eqnarray}
where $h$ is defined by \Eq{defh}.
By using the relations
\begin{eqnarray}
u^{i\prime\prime}_{l_i}u^{f\prime}_{l_f}+u^{f\prime\prime}_{l_f}u^{i\prime}_{l_i}
-u^{i\prime\prime\prime}_{l_i}u^{f}_{l_f}-u^{f\prime\prime\prime}_{l_f} u^i_{l_i} &=&
\left[(u^{i}_{l_i}u^{f}_{l_f})''-2 u^{f\prime\prime}_{l_f} u^{i}_{l_i}-2 u^{f}_{l_f} u^{i\prime\prime}_{l_i}\right]' \\
&=& (2 u^{i\prime}_{l_i}u^{f\prime}_{l_f}-u^{f\prime\prime}_{l_f} u^{i}_{l_i}-u^{f}_{l_f} u^{i\prime\prime}_{l_i})',
\end{eqnarray}
and
\begin{equation}
\beta (u^{f\prime\prime}_{l_f}u^{i}_{l_i}+u^{i\prime\prime}_{l_i}u^{f}_{l_f})=(h+2U)u^{i}_{l_i} u^{f}_{l_f},
\end{equation}
coming from \Eq{Eqrad}, one proves by integrations by parts that
 \begin{eqnarray}\nonumber
\beta \int^b_R G (u^{i\prime\prime}_{l_i}u^{f\prime}_{l_f}+u^{f\prime\prime}_{l_f}u^{i\prime}_{l_i}
-u^{i\prime\prime\prime}_{l_i}u^{f}_{l_f}-u^{f\prime\prime\prime}_{l_f} u^i_{l_i}) d\rho =-\beta \int^b_R G' (u^{i}_{l_i}u^{f}_{l_f})'' d\rho\\ 
+  2\int^b_R G'(h+2U) u^{i}_{l_i}u^{f}_{l_f} d\rho+\left[G [2\beta u^{i\prime}_{l_i}u^{f\prime}_{l_f}-(h+2U)u^{i}_{l_i}u^{f}_{l_f} ]\right]^b_R  \\\nonumber
=-\beta \int^b_R G''' u^{i}_{l_i}u^{f}_{l_f} d\rho  + 2\int^b_R G'(h+2U) u^{i}_{l_i}u^{f}_{l_f} d\rho \\ \label{eqa4}
+\left[\beta [2 G u^{i\prime}_{l_i}u^{f\prime}_{l_f}-G' (u^{i}_{l_i}u^{f}_{l_f})']
+\left[\beta G'' -G(h+2U)\right]u^{i}_{l_i}u^{f}_{l_f} \right]^b_R.
\end{eqnarray}
Let me choose $G=g f(\epsilon,\rho)$ where $\epsilon>0$ and $g$ is a function of class $C^3$ over $(R,\infty)$ and let me take the limit for $b\rightarrow\infty$ in \Eqs{Aeq2} and \eqref{eqa4}.
Every term in \Eqs{Aeq2} and \eqref{eqa4} is assumed to have a finite limit for $b\rightarrow\infty$.
Again, by using a contour integration like in~\Ref{GV71}, it can be proved that the limit for $\epsilon\rightarrow 0$ of the integrals containing a derivative of $f$ at any order is null.
Then, by taking the limit for $\epsilon\rightarrow 0$ of both sides of \Eq{Aeq2} and by using \Eq{eqa4}, \Eq{Eq2} is obtained.
\begin{acknowledgments}
This text presents research results of the interuniversity attraction pole programme P7/12 initiated by the Belgian-state 
Federal Services for Scientific, Technical and Cultural Affairs.
A part of this work was done with the support of the F.R.S.-FNRS. 
TRIUMF receives funding via a contribution through the National Research Council Canada.
\end{acknowledgments}
\end{document}